\documentclass[aps,prl,showpacs,footinbib,superscriptaddress,twocolumn,longbibliography]{revtex4-2}
\usepackage[dvipsnames]{xcolor}
\usepackage{amsfonts}
\usepackage{amsmath}
\usepackage{multirow}
\usepackage{txfonts}
\usepackage{amssymb}
\usepackage{amsbsy}
\usepackage{graphicx}
\usepackage{epstopdf}
\usepackage{color}
\usepackage{braket} 
\usepackage{mathdots} 
\usepackage{booktabs}
\usepackage{indentfirst}
\usepackage{hyperref}
\usepackage{diagbox}
\usepackage{lipsum}
\usepackage{booktabs}
\usepackage{bibentry}
\hypersetup{colorlinks=true, linkcolor=blue, anchorcolor=blue, citecolor=blue,urlcolor=blue}
\usepackage{filecontents}
\usepackage[normalem]{ulem}
\usepackage{comment}

\newcommand{\nocontentsline}[3]{}
\let\origcontentsline\addcontentsline
\newcommand\stoptoc{\let\addcontentsline\nocontentsline}
\newcommand\resumetoc{\let\addcontentsline\origcontentsline}

\makeatletter
\@ifundefined{ifmainmode}{\newif\ifmainmode\mainmodetrue}{}%
\@ifundefined{ifsuppmode}{\newif\ifsuppmode\suppmodetrue}{}%
\@ifundefined{ifcombinedmode}{\newif\ifcombinedmode\combinedmodetrue}{}%
\makeatother

\begin{document}

\author{Yu-Min Hu}
\altaffiliation{These authors contributed equally to this work.}
\affiliation{Max Planck Institute for the Physics of Complex Systems, N\"{o}thnitzer Stra{\ss}e~38, 01187 Dresden, Germany}
\author{Yu-Bo Shi}
\altaffiliation{These authors contributed equally to this work.}
\affiliation{Max Planck Institute for the Physics of Complex Systems, N\"{o}thnitzer Stra{\ss}e~38, 01187 Dresden, Germany}
\affiliation{Department of Physics, National University of Singapore, Singapore 117542, Singapore}
\author{Linhu Li}
\affiliation{Quantum Science Center of Guangdong-Hong Kong-Macao Greater Bay Area, Shenzhen 518048, China}
\author{Gianluca Teza}
\affiliation{Max Planck Institute for the Physics of Complex Systems, N\"{o}thnitzer Stra{\ss}e~38, 01187 Dresden, Germany}

\author{Ching Hua Lee}
\affiliation{Department of Physics, National University of Singapore, Singapore 117542, Singapore}
\author{Roderich Moessner}
\affiliation{Max Planck Institute for the Physics of Complex Systems, N\"{o}thnitzer Stra{\ss}e~38, 01187 Dresden, Germany}

\author{Shu Zhang}
\email{shu.zhang@oist.jp}
\affiliation{Collective Dynamics and Quantum Transport Unit, Okinawa Institute of Science and Technology Graduate University, Onna-son, Okinawa 904-0412, Japan}
\author{Sen Mu}
\email{senmu@pks.mpg.de}
\affiliation{Max Planck Institute for the Physics of Complex Systems, N\"{o}thnitzer Stra{\ss}e~38, 01187 Dresden, Germany}

\date{\today}

\clearpage

\title{Boundary Floquet Control of Bulk non-Hermitian Systems}

\ifmainmode

\begin{abstract}
Boundary perturbations are generally irrelevant for bulk properties in the thermodynamic limit, as they are edge-confined and subextensive. We show that this expectation breaks down in boundary-driven systems exhibiting the non-Hermitian skin effect, where arbitrarily weak boundary Floquet driving reconstructs bulk quasienergy spectra and dynamics. We develop a Floquet non-Bloch band theory that extends generalized Brillouin-zone methods to boundary-driven systems at arbitrary driving frequencies, overcoming the lack of a general framework beyond high-frequency approximations. With representative single- and two-band models, we demonstrate that the boundary driving frequency tunes non-Bloch parity-time symmetry breaking, while its amplitude acts as a finite-size control parameter. Our work establishes boundary Floquet control as a general route for manipulating bulk properties, opening a new avenue for dynamical engineering in driven open systems.
\end{abstract}

\maketitle

\emph{Introduction.}|Floquet driving has become a central route for controlling classical and quantum systems because it combines rich nonequilibrium dynamics with strong engineering flexibility \cite{oka2019floquet, eckardt2017colloquium, rudner2020band}. Examples include the realization of Floquet topological phases of matter~\cite{lindner2011floquet, Kitagawa2010topological,lee2018floquet,Rudner2013Anomalous, cayssol2013floquet,li2018realistic}, the design of controllable effective Hamiltonians~\cite{bukov2015universal, goldman2014periodically,stegmaier2025topological}, stabilization of long-lived prethermal dynamical regimes~\cite{Abanin2015Exponentially, Abanin2017effective, Else2017prethermal, Rubio2020Floquet, ho2023quantum} and discrete time crystals~\cite{Khemani2016Phase, Else2016Floquet, zhang2017observation, khemani2019briefhistorytimecrystals, else2020discrete,Zaletel2023dtc}, which have been implemented across platforms ranging from solid-state materials~\cite{aeschlimann2021survival,wang2013observation,merboldt2025observation} to various quantum simulators~\cite{zhang2022digital, jotzu2014experimental, frey2022realization, nguyen2024programmable, chen2023robust, katz2025floquet, shen2024enhanced,Koyluoglu2025floquet}. 
Beyond global bulk modulation,  local driving--especially  boundary driving--is emerging as a powerful approach, with demonstrated advances in tuning topological edge states~\cite{fedorova2019limits, Mukherjee2024emergent}, navigating dynamics and entanglement of Hermitian Hamiltonians at criticality~\cite{Berdanier2017floquet, lin2025local}, probing transport in open quantum systems~\cite{Prosen2011exact, Prosen2011Open, Landi2022nonequilibrium}, enabling anomalous relaxation phenomena in classical and quantum settings~\cite{Teza2023_mpemba_boundary,teza2026speedups} and designing quantum circuits~\cite{Chiara2025integrability}.

Non-Hermitian systems offer another powerful route to nonequilibrium phase design and control \cite{Ashida2021, Bergholtz2021RMP, okuma2023non}. A hallmark is the non-Hermitian skin effect~(NHSE)~\cite{yao2018edge,  lee2019anatomy, kunst2018biorthogonal, Lin_2023, ding2022non}, where, under open boundary conditions (OBC), extensive eigenstates localize at boundaries, corresponding to non-trivial point-gap topology under periodic boundary conditions (PBC) \cite{Okuma2020topological, zhang2020correspondence,yang2022designing,wang2024amoeba,zhang2022universal}. It underlies numerous key phenomena of non-Hermitian systems, including unconventional non-Hermitian bulk-boundary correspondence \cite{Yao2018Chern,ghatak2020observation,xiao2020non,helbig2020generalized,weidemann2020topological,zhang2021tidal,li2019geometric}, anomalous non-Hermitian dynamics \cite{Song2019chiral,Haga2021Liouvillian,Longhi2019probing, Yang2024realtime, xue2025nonbloch,Longhi2022healing, Xue2022non,gu2022transient,Longhi2019nonBloch,Xiao2021observation,Hu2024Geometric,qin2024kinked,li2024observation,Qin2024occupation,yang2025reversing,yang2025beyond}, and promising applications such as non-Hermitian amplification and sensing via strong spectral response \cite{wanjura2020topological,xue2021simple,Budich2020non,Florian2022quantum,li2021quantized,Liang2022Anomalous}. Such open-boundary localized eigenstates (skin modes) are typically stable against static boundary perturbations, since their eigenvalues and localization lengths are set by the bulk properties, as described by non-Bloch band theory and the generalized Brillouin zone (GBZ)~\cite{yao2018edge, Yokomizo2019non, Yang2020non}.

In general,  because boundary terms, whether static or time-dependent, are subextensive, one may expect them to be negligible for bulk-determined quantities in the thermodynamic limit, as is the case for a gapped Hermitian system. Indeed, studies on Floquet non-Hermitian systems have so far mainly focused on bulk engineering in the high-frequency regime, with an emphasis on effective-band descriptions and topological characterizations~\cite{Zhou2018non,Zhang2020nonhermitan, Wu2020floquet, Cao2021nonhermitian, park2022revealing, Liu2022symmetry, Sun2024photonic, Ji2024Generalized, Lin2024observation, shi2024Floquet, Tong2025observation, Roy2025Topological, koch2025liouvillian, wu2025breakdownnon, Zhou2025non,zhou2026topology}. 
However, the presence of the NHSE can fundamentally overturn the usual expectation and open a striking possibility for the boundary drive to act nonperturbatively on the bulk. 

In this Letter, we show that a time-periodic drive applied only at the boundaries can resonantly hybridize Floquet replicas of skin modes and thereby induce a global reconstruction of bulk spectral and dynamical properties. This establishes boundary Floquet driving as a generic mechanism for controlling bulk non-Hermitian systems. Since this effect is rooted in Floquet-zone mixing, its description demands a boundary-Floquet framework beyond the high-frequency regime. We fill this theoretical gap by developing a Floquet non-Bloch band theory valid at arbitrary boundary driving frequency. In particular, our framework provides a geometric description of non-Hermitian Floquet-zone mixing in the low-frequency regime. It directly captures the OBC quasienergy spectrum in the thermodynamic limit without relying on any truncation to an effective Floquet Hamiltonian. We further show that, in the thermodynamic limit, even an infinitesimal boundary drive makes the driving frequency a powerful knob for tuning parity-time symmetry breaking, spectral structure, and dynamical response. Boundary Floquet driving therefore offers a practical route to controlling non-Hermitian dynamics.

\begin{figure*}[t]
	\centering
	\includegraphics[width=1.0\linewidth]{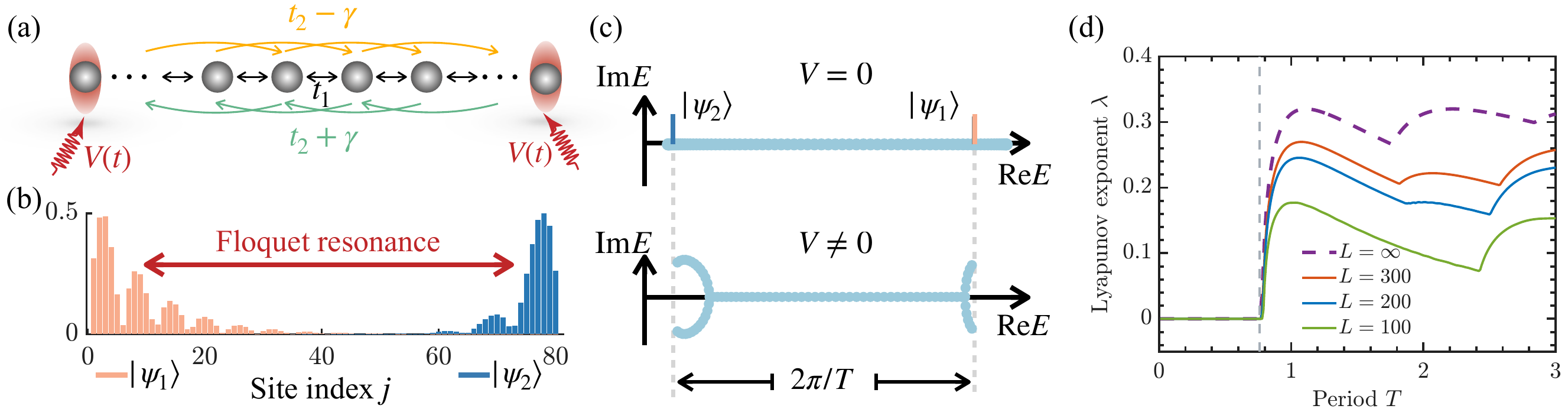}
\caption{Key mechanism underlying the boundary Floquet control of bulk systems with the non-Hermitian skin effect. (a) Schematic illustration of the boundary driving protocol for the time-periodic Hamiltonian $H(t)=H_0+V(t)$, with the static bulk Hamiltonian $H_0$ in Eq.~\eqref{eq:H0} and the boundary drive $V(t)$ with period duration $T$ in Eq.~\eqref{eq:boundary_step}.  (b) An example of two skin modes of the static Hamiltonian $H_0$ whose energies differ by $2\pi/T$, labeled in the upper panel of (c). They exhibit distinct localization behaviors and are resonantly coupled via $V(t)$. (c) Upper panel: OBC spectrum of the static Hamiltonian $H_0$ (i.e., $V(t)=0$). Lower panel: OBC quasienergy spectrum of $H(t)$ with $V(t)\neq 0$. We have set $V=0.01$ in the lower panel; other parameters are $t_1=2$, $t_2=0.15$, $\gamma=0.16$, $T=0.9$, and $L=80$.  (d) Lyapunov exponent $\lambda=\lim_{t\to\infty}[\ln\braket{\psi(t)|\psi(t)}]/(2t)$ as a dynamical signature during the evolution for an initial state $\ket{\psi(0)}=\ket{L/2}$, 
corresponding to the largest imaginary part of the OBC quasienergy spectrum. Solid colored lines denote finite-size results for different system sizes, while the dashed purple line shows the thermodynamic-limit prediction from our Floquet non-Bloch band theory. Our theory also predicts the threshold value $T_c$ for the Floquet-induced parity-time symmetry breaking, shown as the dashed gray line in (d).}
	\label{fig1}
\end{figure*}

\emph{Boundary Floquet protocol.}| Our Floquet non-Bloch band theory applies generally to non-Hermitian systems subject to time-periodic boundary driving, $H(t)=H_0+V(t)$, where $V(t+T)=V(t)$ denotes time-periodic boundary terms with period $T$. Its dynamics is governed by the Floquet operator
\begin{equation}\label{eq:U_F}
U_{\mathrm{F}}=\mathcal{T}\exp\!\left(-i\int_{0}^{T} H(t)\,\mathrm{d}t\right)
\equiv\exp(-i H_{\mathrm{F}} T),
\end{equation}
where $\mathcal{T}$ denotes time ordering and $H_{\mathrm{F}}$ is the effective time-independent Floquet Hamiltonian. The eigenstates and quasienergy spectrum of $H_{\mathrm{F}}$ provide the central characterization of the Floquet dynamics. In the high-frequency limit ($T \rightarrow 0$), one has $\lim_{T \rightarrow 0} H_{\mathrm{F}} = (1/T)\int_{0}^{T} H(t)\,\mathrm{d}t=H_0$. The dynamics can thus be described by the time-averaged Hamiltonian, namely the zeroth order term in the Floquet-Magnus expansion. Upon lowering the driving frequency, higher-order expansion terms generated by the commutator $[H(t), H(t')]$ near the boundaries become more prominent and introduce frequency-dependent corrections to $H_F$. 

In this work, we go beyond the high-frequency expansion to treat the regime where strong Floquet-zone mixing requires the development of a genuinely Floquet non-Bloch description. Our theoretical framework applies to generic boundary-driven one-dimensional multiband systems, where the static bulk Hamiltonian is of the non-Bloch form $h(\beta)=\sum_{n=-m}^{m}h_n\beta^n$, with $q$ orbitals per unit cell, maximum hopping range $m$, and $q\times q$ matrices $h_n$ specifying the intercell hopping elements. More broadly, our theory also applies to a large class of periodically driven non-Hermitian systems where the Floquet-zone mixing is dominated by the boundary terms of Floquet operators.  Multiband models and bulk-driven systems are discussed in the Supplemental Material (SM)~\cite{supp_mat}.

Here, we demonstrate the main mechanism of our theory with a minimal model where non-trivial boundary-driving-induced effects occur, taking $q=1$ and $m=2$ in the generic form above.  As shown in Fig.~\ref{fig1}(a), we consider a static bulk Hamiltonian $H_0$:
\begin{equation}
    \begin{aligned}
            H_0{=}&\sum_{j=1}^{L-1}t_1(c_j^\dagger c_{j+1} {+ }c_{j+1}^\dagger c_j){+}\sum_{j=1}^{L-2}(t_2{+}\gamma)c_j^\dagger c_{j+2} {+}(t_2{-}\gamma)c_{j+2}^\dagger c_j,
    \end{aligned}\label{eq:H0}
\end{equation}
where $c_j$ and $c_{j}^\dagger$ are particle annihilation and creation operators on the $j$th site, $L$ is the system size, $t_1$ and $t_2$ are Hermitian hopping amplitudes, and $\gamma$ sets the amplitude of non-Hermitian hoppings between next-nearest neighbors.
Its OBC spectrum follows from the standard non-Bloch band theory of static non-Hermitian systems by evaluating a non-Bloch Hamiltonian $h(\beta)$ along the GBZ~\cite{yao2018edge, Yokomizo2019non}, which is given by
\begin{equation}\label{eq:non-Bloch}
    h(\beta)=(t_2+\gamma)\beta^2+(t_2-\gamma)\beta^{-2}+t_1(\beta+\beta^{-1}), \quad \beta \in \mathbb{C}.
\end{equation} 
For $\beta = e^{ik}$ with real momentum $k$, Eq.~\eqref{eq:non-Bloch} reduces to the conventional Bloch Hamiltonian under PBC. 
A time-dependent potential $V(t)$ with period duration $T$ (frequency $2\pi/T$) is applied at the two ends: $V(t+T)=V(t)$ and
\begin{equation}\label{eq:boundary_step}
    V(t)=\begin{cases}
        +V(c_1^\dagger c_1+c_L^\dagger c_L), \quad &0\le t<T/2;\\
       - V(c_1^\dagger c_1+c_L^\dagger c_L), \quad &T/2\le t< T.
    \end{cases}
\end{equation}

\emph{Floquet-zone folding mechanism.}|As the driving frequency drops below the energy bandwidth of $H_0$, different Floquet replicas begin to overlap within the quasienergy Brillouin zone. The resulting Floquet-zone folding enables the frequency-dependent terms to resonantly couple skin modes of $H_0$ whose energies differ by integer multiples of $2\pi/T$, as schematically illustrated in Fig.~\ref{fig1}(b) and (c). This in turn allows skin modes with different localization lengths to hybridize, drastically reshaping the quasienergies and eigenmodes and rendering them strongly dependent on the driving period $T$. Though this spectral reshaping due to coupling between skin modes is reminiscent of the critical NHSE in static systems~\cite{li2020critical, Yokomizo2021scaling,song2024fragile, shu2024ultraspectral, cheng2025stochasticity,Yi2025critical,liu2025non,cheng2025stochasticity,qin2025many}, the setting is fundamentally different: we consider here the Floquet hybridization caused by boundary  driving in a non-Hermitian system.

For illustration, consider two eigenmodes of the static Hamiltonian $H_0$ under OBC, e.g. see Fig.~\ref{fig1}(b), $\psi_1(j)\sim e^{\kappa_1 j}$ and $\psi_2(j)\sim e^{\kappa_2 j}$ with eigenenergies $E_1$ and $E_2$ separated by $2\pi/T$ and inverse localization lengths $\kappa_{\alpha}$ ($\alpha = 1,2$) respectively. Note that $\kappa_\alpha>0$ ($\kappa_\alpha<0$) indicates localization towards the right (left) boundary. The corresponding left eigenmodes, defined as $\bra{\tilde\psi_\alpha}H_0=\bra{\tilde\psi_\alpha}E$, are localized in the opposite direction $\tilde{\psi}_\alpha (j)\sim e^{-\kappa_\alpha j}$. 
The boundary driving resonantly couples these two modes. {Applying degenerate perturbation theory gives two off-diagonal matrix elements $\chi_1 = \braket{\tilde\psi_1|V_{2\pi/T}|\psi_2}\sim V[e^{(\kappa_2-\kappa_1)L}+e^{\kappa_2-\kappa_1}]$ and $\chi_2 =\braket{\tilde\psi_2|V_{2\pi/T}|\psi_1}\sim V[e^{(\kappa_1-\kappa_2)L}+e^{\kappa_1-\kappa_2}]$ with $V_\omega=\int_0^T (\mathrm{d}\omega/2\pi) e^{i\omega t} \,V(t)$, whereas the diagonal parts remains $O(V)$ with a negligible size dependence. Whenever $\kappa_1\ne\kappa_2$, the hybridization induces an exponentially large energy split $\sqrt{\chi_1 \chi_2} \sim V e^{|\kappa_1-\kappa_2|L/2}$. Consequently, even a weak boundary drive can trigger a global reconstruction of the quasienergy spectrum that deviates markedly from that of the static Hamiltonian $H_0$ [Fig.~\ref{fig1}(c)]. Note that in the special case where $\kappa_1=\kappa_2$ for any two eigenmodes, i.e. a circular GBZ, a weak boundary Floquet drive remains perturbative~\footnote{The perturbation also remains small in static 1D systems with time-independent weak boundary perturbations. Apart from a few special cases~\cite{song2024fragile}, the finite-size level spacing in generic 1D systems scales as $E_2-E_1\propto 1/L$, along with $|\kappa_2-\kappa_1|\propto 1/L$. As a result, the boundary-induced coupling remains finite since $|\kappa_2-\kappa_1|L\sim O(1)$.}.

Indeed, we find that a weak boundary drive triggers a real-to-complex spectral transition, corresponding to the breaking of non-Bloch parity-time (PT) symmetry~\cite{Longhi2019probing,Longhi2019nonBloch,Xiao2021observation,Hu2024Geometric}.  This symmetry, which originates from $H(t)=[H(t)]^*$, ensures that the OBC quasienergy spectrum is either entirely real or consists of complex-conjugate pairs, even though the PBC spectrum is always complex for any $\gamma\neq 0$.  As the maximal imaginary part of the OBC quasienergy spectrum controls the long-time dynamics~\cite{Longhi2019probing, Yang2024realtime, xue2025nonbloch},  such a real-to-complex spectral transition (i.e., PT transition) leads to pronounced dynamical implications, which can serve as an experimental signature in non-Hermitian platforms~\cite{Xiao2021observation}. Consider the non-Hermitian Floquet dynamics $i\partial_t\ket{\psi(t)}=H(t)\ket{\psi(t)}$ starting from an initial state $\ket{\psi(0)}=\ket{L/2}$. Tracking the wavefunction norm at long time yields the Lyapunov exponent $\lambda=\lim_{t\to\infty} \left[\ln \braket{\psi(t)|\psi(t)}/2t\right]$, which is set by the spectrum’s maximal imaginary part. As the driving period $T$ exceeds a threshold value $T_c$, the Lyapunov exponent exhibits a sharp transition from zero to a finite value [see Fig.~\ref{fig1}(d)], indicating a real-to-complex spectral transition. 
The strong finite-size dependence and the Floquet GBZ theory that predicts the $L \rightarrow \infty$ limit will be discussed below.

\emph{PT transition and emergent finite-size dependence.}|
We characterize the PT transition and the associated finite-size effect by defining a ratio $\eta \in[0,1]$ as the fraction of complex eigenvalues in the OBC quasienergy spectrum. Our numerical results are presented as color maps in Fig.~\ref{fig2} with $\eta=0$ ($\eta>0$) indicating a real (complex) quasienergy spectrum. For short $T$, as in Fig.~\ref{fig2}(a), the spectrum remains entirely real and nearly identical to that of the static Hamiltonian $H_0$, since the boundary drive is off-resonant and weakly affects the system. For $T>T_c$, however, the boundary drive induces inter-zone hybridization between skin modes, leading to complex quasienergies and a nonzero $\eta$ with pronounced system-size dependence.

In addition to the Floquet-induced PT transition, increasing the non-Hermitian strength $\gamma$ can also drive a spectral transition. This transition [horizontal cyan line in Fig.~\ref{fig2}(b)] can be understood geometrically as the emergence of cusps on the static GBZ of $H_0$ \cite{Hu2024Geometric}, which in turn also gives the critical value $\gamma_c$. As we will show later, the Floquet-induced PT transition [nearly-vertical gray lines in Fig.~\ref{fig2}(a,b)] embodies a similar geometric structure once we properly construct the Floquet GBZ, which provides a framework to determine the PT transition phase boundary.

\begin{figure}[t]
    \centering
    \includegraphics[width=8cm]{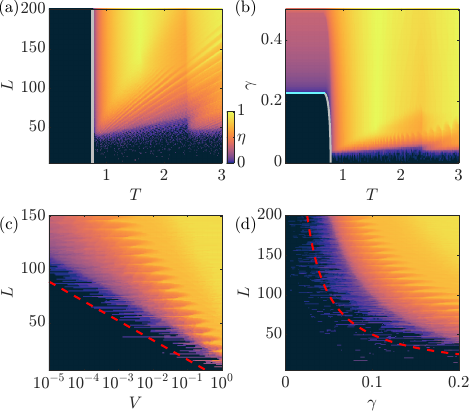}
    \caption{Phase diagrams of the ratio $\eta$ as the fraction of OBC quasienergies that are complex, showing the Floquet-induced PT-symmetry breaking and its pronounced finite-size dependence. (a) The phase diagram of the system size $L$ and driving period $T$, with $\gamma=0.16$ and $V=0.01$. (b) The phase diagram of the non-Hermitian strength $\gamma$ and driving period $T$, with $L=200$ and $V=0.01$.  (c) The phase diagram of the system size $L$ and driving strength $V$, with $T=2$ and $\gamma=0.16$. (d) The phase diagram of the system size $L$ and non-Hermitian strength $\gamma$, with $T=2$ and $V=0.01$. In all plots, we set $t_1=2$ and $t_2=0.15$. The solid lines in (a) and (b) are theoretical predictions from our Floquet GBZ theory. The red dashed lines indicate the scaling relation in Eq. \eqref{eq:scaling}: $L_c\propto -\log V$ in (c) and $L_c\propto 1/\gamma$ in (d), revealing that even weak boundary driving can trigger a bulk spectral transition, only beyond a parametrically large system size.}
    \label{fig2}
\end{figure}

\begin{figure*}[t]
	\centering
	\includegraphics[width=1.0\textwidth]{
    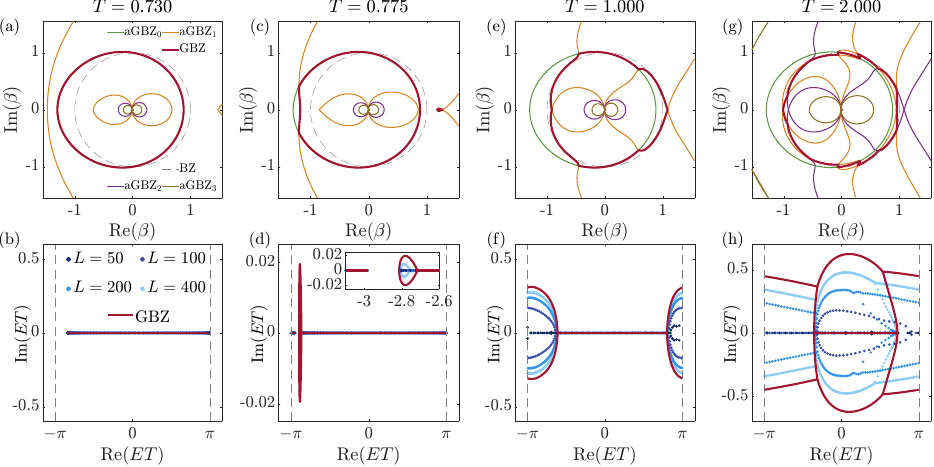}
	\caption{The Floquet GBZ and aGBZs (top row), and the corresponding OBC quasienergy spectra (bottom row), for representative driving periods $T$ with $\gamma=0.16$, $t_1=2$, $t_2=0.15$, and $V=0.01$. Here ${\rm aGBZ}_{\ell}$ denotes the auxiliary GBZ associated with the $\ell$th Floquet replica $E\pm2\pi \ell/T$. As $T$ increases, neighboring Floquet replicas approach and hybridize, which is captured geometrically by the evolution of these curves in the top row. In the bottom row, colored points are obtained by exactly diagonalizing the OBC Floquet operator $U_{\text{F}}$ for different system sizes, while solid lines are the thermodynamic-limit predictions from our Floquet GBZ theory. This shows that Floquet-zone hybridization in the large-$T$ regime gives rise to cusps in the Floquet GBZ from which complex spectral branches emerge.}
	\label{fig3}
\end{figure*}

Before diving into our Floquet GBZ theory, we provide further insight into the finite-size effects of complex spectra as shown in Figs.~\ref{fig2}(c) and (d). Recall the illustrative picture of skin-mode coupling presented in Fig.~\ref{fig1}(b): The hybridization strength is given by $V e^{|\kappa_1-\kappa_2|L/2}$. Meanwhile, for small $\gamma$, static non-Bloch band theory shows that the GBZ deviates from the conventional Brillouin zone by $O(\gamma)$, so that $|\kappa_1-\kappa_2| \propto \gamma$ for two generic skin modes at different energies. We can therefore identify a characteristic length scale 
\begin{equation}\label{eq:scaling}
L_c \propto -(\log V)/\gamma.
\end{equation}
For system sizes below $L_c$, the skin-mode coupling remains perturbative. This system-size dependence accounts for the regions with nearly-zero $\eta$ in the low-driving-frequency regime in Fig.~\ref{fig2}(a) and (b). As plotted by red dashed curves in Figs.~\ref{fig2}(c) and \ref{fig2}(d), the OBC quasienergy spectrum becomes noticeably complex when the system size exceeds $L_c$: $L_c\propto -\log V$ for weak boundary-driving strength $V$, and $L_c\propto 1/\gamma$ for weak non-Hermitian hopping amplitude $\gamma$. It also suggests that in the thermodynamic limit ($L\to\infty$), even an infinitesimal boundary drive can nonperturbatively induce a sharp spectral change.

\emph{Geometric perspective from Floquet GBZ theory.}|We now develop a Floquet non-Bloch band theory that can quantitatively capture the OBC quasienergy spectrum across the full range of driving frequency of boundary drives.  The central step is to construct a Floquet GBZ. This Floquet GBZ provides a geometric characterization of the boundary-driven PT transition due to hybridization of skin modes originating from different Floquet zones, and allows for identifying the corresponding critical period $T_c$ in the thermodynamic limit.

We recall that, for a static non-Hermitian system, the OBC spectrum is obtained from solving the characteristic equation $f(\beta,E)=\det[h(\beta)-E]=0$. Under periodic driving, by contrast, the quasienergy is defined only modulo the driving frequency, so one must in principle consider the full hierarchy of characteristic equations $f(\beta,E+2\pi\ell/T)=0$ with $\ell\in\mathbb{Z}$, corresponding to all Floquet replicas. In the high-frequency limit, however, the shifted replicas with $\ell\neq 0$ are well separated from $E$ and have negligible effect, so the Floquet GBZ reduces to the static GBZ. At fixed $E$, one then orders the $2M$ roots of $f(\beta,E)=0$ as $|\beta_1(E)|\le\cdots\le|\beta_{2M}(E)|$, with $M=mq$, and determines the GBZ from the equal-modulus condition of middle two roots: $|\beta_M(E)|=|\beta_{M+1}(E)|$, shown as the red curve in Fig.~\ref{fig3}(a). The OBC spectrum then follows by evaluating $h(\beta)$ along this curve in the complex $\beta$ plane~\cite{yao2018edge, Yokomizo2019non}.

To treat the low-frequency regime, it is useful to first introduce an intermediate object, the Floquet auxiliary generalized Brillouin zone (aGBZ). It is defined by the equal-modulus condition between roots belonging to different Floquet zones,
$\text{aGBZ}_{\ell}\equiv\{\beta: |\beta_i(E)|=|\beta_j(E\pm 2\pi\ell/T)| \}$,
where $\ell=0,1,2,\cdots$ labels the Floquet zone shifted by $\pm2\pi\ell/T$. These curves can be evaluated by the resultant method introduced in Ref.~\cite{Yang2020non}, with details given in the SM~\cite{supp_mat}.
In the high-frequency regime, the Floquet GBZ is a subset of $\mathrm{aGBZ}_0$, reflecting that the OBC spectrum is controlled by the central Floquet sector and the shifted replicas remain effectively decoupled. 
This is seen geometrically in Fig.~\ref{fig3}(a): the red GBZ loop lies entirely on $\mathrm{aGBZ}_0$, while other auxiliary curves $\mathrm{aGBZ}_{\ell\neq 0}$, such as the orange loops, stay well separated from it.
Since these curves do not meet, Floquet-zone hybridization is absent and the Floquet GBZ reduces to the static GBZ.

The key change occurs at lower driving frequency, when $\mathrm{aGBZ}_1$ moves toward and eventually intersects the high-frequency Floquet GBZ. This signals that roots from different Floquet zones exchange their modulus ordering, so the equations $f(\beta,E+2\pi\ell/T)=0$ in the sectors $\ell=-1,0,1$ must now be treated on equal footing. Sorting the resulting $6M$ roots as $|\tilde{\beta}_1(E)| \le \cdots \le |\tilde{\beta}_{6M}(E)|$, the Floquet GBZ is fixed by the equal-modulus condition for the middle pair $|\tilde{\beta}_{3M}(E)| = |\tilde{\beta}_{3M+1}(E)|$. Geometrically, as $T$ increases, the intersection with ${\rm aGBZ}_1$ inserts new segments and cusp points into the Floquet GBZ, leading directly to a reconstruction of the OBC quasienergy spectrum~\cite{Hu2024Geometric}, as shown in Figs.~\ref{fig3}(c-f). The critical period $T_c$ of PT-symmetry breaking is thus identified when ${\rm aGBZ}_1$ first touches the static GBZ. The phase boundaries obtained in this way are shown by the gray lines in Figs.~\ref{fig2}(a) and (b).

Upon further lowering the frequency, additional $\mathrm{aGBZ}_\ell$ curves enter and intersect the Floquet GBZ, producing more cusps and segments. We denote by $\ell_0$ the largest index of $\mathrm{aGBZ}_\ell$ involved and collect the roots of $f(\beta,E+2\pi\ell/T)=0$ with $\ell = -\ell_0, \ldots, \ell_0$ into one sequence $|\tilde{\beta}_1(E)| \le \cdots \le |\tilde{\beta}_{2(2\ell_0+1)M}(E)|$. The Floquet GBZ is again determined by the middle pair, $|\tilde{\beta}_{(2\ell_0+1)M}(E)| = |\tilde{\beta}_{(2\ell_0+1)M+1}(E)|$. 
The case of $\ell_0 = 2$ is shown in Fig.~\ref{fig3}(g-h), where the extra intersections generate additional GBZ segments and new branches in the OBC quasienergy spectrum. In practice, without a prior knowledge of $\ell_0$, the construction is implemented with a cutoff $\ell_c$ and justified by convergence as $\ell_c$ increases \cite{supp_mat}. Nevertheless, the resulting Floquet GBZ is an intrinsic model-determined geometric object, irrelevant to $\ell_c$. We note that the finite-size quasienergies in the lower row of Fig.~\ref{fig3}, as well as the finite-size Lyapunov exponents in Fig. \ref{fig1}(d), converge to our Floquet GBZ prediction as system size increases, providing quantitative confirmation of our theory.

\textit{Discussion.}|We have demonstrated boundary Floquet control for bulk non-Hermitian systems, where tuning the boundary driving frequency controls the quasienergy spectrum and Floquet dynamics. We have also established a Floquet non-Bloch band theory valid at arbitrary boundary driving frequency, providing a unified framework for determining the quasienergy spectra and critical frequencies for spectral transitions in the thermodynamic limit. Our theory is constructed entirely from the bulk Hamiltonian. This is because Floquet-zone folding is governed solely by the driving frequency and the boundary drive leaves the bulk hoppings unchanged. Therefore, the spectra in the thermodynamic limit are independent of the details of the boundary drive.

To illustrate the generality of our theory, we present a representative boundary-driven two-band model in the SM~\cite{supp_mat}. Our theory also extends to a large class of bulk-driven non-Hermitian systems, provided that the time-periodic non-Bloch Hamiltonian $h(\beta,t)$ is mutually commuting at different times, $[h(\beta,t),h(\beta,t^\prime)]=0$. The corresponding commutator $[H(t), H(t^\prime)]$ of the OBC Hamiltonian $H(t)$ does not vanish only near the boundaries. We provide a detailed analysis of this class with a bulk-driven example in the SM \cite{supp_mat}.

Our framework opens several new directions for investigating nonequilibrium phenomena in Floquet non-Hermitian systems. Natural directions for future work include extensions to higher dimensions and Liouvillian dynamics, and the investigation of topological characterization of boundary-Floquet non-Hermitian phases and associated phenomena. The predicted effects, such as Floquet-induced non-Bloch PT symmetry breaking, are well-suited for realization and observation in a broad range of state-of-the-art non-Hermitian platforms.

\textit{Acknowledgments.---} We thank Zhong Wang, Alexander Poddubny, Jan Carl Budich, Masudul Haque, and Marin Bukov for insightful discussions and comments. 
This work was in part supported by the Deutsche Forschungsgemeinschaft under grant cluster of excellence ctd.qmat (EXC 2147, project-id 390858490). 
L. Li acknowledges support from the National Natural Science Foundation of China (Grant No. 12474159) and the Guangdong Provincial Quantum Science Strategic Initiative (Grants No. GDZX2504003 and GDZX2504006). 
C.H. Lee and Y.-B. Shi acknowledges support from the Singapore Ministry of Education Tier II. grants MOE-T2EP50224-0007 and MOE-T2EP50224-0021 (WBS nos. A-8003505-01-00 and A-8003910-00-00).

\emph{Note added.--} We are aware of a related work  by Bo Li {\it et al} \cite{Li2026boundary}. 

\bibliography{references}

\end{document}


\author{Yu-Min Hu}
\altaffiliation{These authors contributed equally to this work.}
\affiliation{Max Planck Institute for the Physics of Complex Systems, N\"{o}thnitzer Stra{\ss}e~38, 01187 Dresden, Germany}
\author{Yu-Bo Shi}
\altaffiliation{These authors contributed equally to this work.}
\affiliation{Max Planck Institute for the Physics of Complex Systems, N\"{o}thnitzer Stra{\ss}e~38, 01187 Dresden, Germany}
\affiliation{Department of Physics, National University of Singapore, Singapore 117542, Singapore}
\author{Linhu Li}
\affiliation{Quantum Science Center of Guangdong-Hong Kong-Macao Greater Bay Area, Shenzhen 518048, China}
\author{Gianluca Teza}
\affiliation{Max Planck Institute for the Physics of Complex Systems, N\"{o}thnitzer Stra{\ss}e~38, 01187 Dresden, Germany}

\author{Ching Hua Lee}
\affiliation{Department of Physics, National University of Singapore, Singapore 117542, Singapore}
\author{Roderich Moessner}
\affiliation{Max Planck Institute for the Physics of Complex Systems, N\"{o}thnitzer Stra{\ss}e~38, 01187 Dresden, Germany}

\author{Shu Zhang}
\email{shu.zhang@oist.jp}
\affiliation{Collective Dynamics and Quantum Transport Unit, Okinawa Institute of Science and Technology Graduate University, Onna-son, Okinawa 904-0412, Japan}
\author{Sen Mu}
\email{senmu@pks.mpg.de}
\affiliation{Max Planck Institute for the Physics of Complex Systems, N\"{o}thnitzer Stra{\ss}e~38, 01187 Dresden, Germany}

\date{\today}

\clearpage

\title{Supplemental material for Boundary Floquet Control of Bulk non-Hermitian Systems}

\ifmainmode

\maketitle

\tableofcontents

\section{General formula for Floquet non-Bloch band theory}\label{sec:theory}
In the main text, we developed the Floquet GBZ theory for predicting quasienergy spectra in non-Hermitian systems with boundary drives. In this supplemental material, we first formulate the Floquet non-Bloch band theory in its most general form for Floquet non-Hermitian systems and then present a practical procedure for its implementation. Although the discussion in the main text is illustrated using a boundary-driven non-Hermitian model with a static bulk Hamiltonian, we show here that the theory also applies to a broad class of bulk-driven non-Hermitian systems.

\subsection{The general applicability}\label{sec:general}
We consider a 1D time-periodic multiband non-Hermitian Hamiltonian $H(t)=H(t+T)$ under open boundary conditions, where $T$ is the driving period. It has two parts: 
\begin{equation}\label{seq:H(t)}
    H(t)=H_{\text{bulk}}(t)+H_{\text{edge}}(t).
\end{equation} 
$H_{\text{edge}}(t)$ describes the boundary Floquet drives, whose action is restricted to two boundary regimes (e.g., Eq. (3) in the main text).  Besides, we also consider a generic time-periodic short-ranged bulk Hamiltonian $H_{\text{bulk}}(t)$, with a corresponding time-dependent non-Bloch Hamiltonian 
\begin{eqnarray}\label{seq:hbetat}
    h(\beta,t)=\sum_{n=-m}^mh_n(t)\beta^n.
\end{eqnarray} 
Here, $m$ is the largest hopping range. With $q$ being the number of orbitals in each unit cell, the $q\times q$ matrices $h_n(t)$ are time-dependent hopping elements between different unit cells. The illustrative example in the main text has a time-independent bulk Hamiltonian with $m=2$ and $q=1$. An additional example with bulk drives will be presented shortly in Sec.  \ref{sec:bulkdriven}. 

We focus on the OBC quasienergy spectra for the Floquet operator $U_{\text{F}}$, which is defined as
\begin{eqnarray}\label{seq:UF}
     U_{\text{F}}\equiv\mathcal{T}\exp[-i\int_0^TH(t)\mathrm{d}t]=\exp(-iH_{\text{F}}T),
\end{eqnarray}
where $\mathcal{T}$ is the time-ordering operator and $H_{\text{F}}$ is the effective Floquet Hamiltonian under open boundary conditions. 

In the main text, we have developed a Floquet GBZ theory that faithfully predicts the OBC spectral information of non-Hermitian systems with a static bulk Hamiltonian [i.e., a time-independent $H_{\text{bulk}}(t)= H_{\text{bulk}}$] and boundary Floquet drive. This is achieved by solving some algebraic equations of a time-independent bulk non-Bloch Hamiltonian [e.g., see $h(\beta)$ in Eq. (4) of the main text]. In this case, the driving frequency of the boundary Floquet process plays an indispensable role in Floquet zone folding and skin-mode hybridization, whereas the microscopic details of $H_{\text{edge}}(t)$ are irrelevant for determining the bulk OBC quasienergy spectra. 
 
In this supplemental material, we discuss the general applicability of Floquet GBZ theory. In other words, we will discuss when and how the OBC quasienergy spectra of 1D Floquet non-Hermitian systems can be efficiently extracted by simply solving some algebraic equations related to a (time-dependent) bulk non-Bloch Hamiltonian $h(\beta,t)$. To this end, we define the Floquet operator in reciprocal space 
\begin{equation}
    u_{\text{F}}(\beta)\equiv\mathcal{T}\exp[-i\int_0^Th(\beta,t)\mathrm{d}t]=\exp[-ih_{\text{F}}(\beta)T] ,
\end{equation}where $h_{\text{F}}(\beta)$ is the Floquet non-Bloch Hamiltonian. Similar to the conventional non-Bloch band theory in static non-Hermitian systems \cite{yao2018edge,Yokomizo2019non},  identifying the OBC eigenvalues $\lambda=e^{-iET}$ of $U_{\text{F}}$ in Eq.  \eqref{seq:UF} with the quasienergy $E$ requires solving the characteristic equation $\det[u_{\text{F}}(\beta)-\lambda]=0$, or equivalently, solving an infinite set of equations 
\begin{equation}\label{seq:characteristic}
f(\beta,E+2\ell\pi/T)=\det[h_{\text{F}}(\beta)-E- 2\pi\ell/T]=0,\quad \ell\in\mathbb{Z}.
\end{equation}
For generic Floquet non-Hermitian systems, $h_{\text{F}}(\beta)$ is an infinite Laurent series, encoding long-range \emph{bulk} hopping terms in $H_{\text{F}}$ stemming from $[h(\beta,t),h(\beta,t^\prime)]\ne0$ for some $t\ne t^\prime$.  This fact makes it generally hard to solve these characteristic equations. Therefore, to facilitate the development of a faithful Floquet non-Bloch band theory, we hereafter focus on the situation
\begin{equation}
    [h(\beta,t),h(\beta,t^\prime)]=0.
    \label{seq:non-Bloch_commutator}
\end{equation} 
This commutation condition implies a simple expression of the Floquet non-Bloch Hamiltonian
\begin{equation}\label{seq:h_F_beta}
    h_{\text{F}}(\beta)=\frac{1}{T}\int_0^Th(\beta,t)\mathrm{d}t=\sum_{n=-m}^m\bar{h}_n\beta^m.
\end{equation}
Here, we define $\bar h_n=\frac{1}{T}\int_0^Th_n(t)\mathrm{d}t$ for $h_n(t)$ in Eq.  \eqref{seq:hbetat}. Since $h(\beta,t)$ in Eq. \eqref{seq:hbetat}, which encodes the short-ranged bulk Hamiltonian $H_{\text{bulk}}(t)$, is a Laurent polynomial with finitely many powers of $\beta$, the resulting Floquet non-Bloch Hamiltonian $h_{\mathrm{F}}(\beta)$ is likewise a finite Laurent polynomial.

The commutator in Eq.  \eqref{seq:non-Bloch_commutator} indicates that $H_{\text{bulk}}(t)$ under periodic boundary conditions, denoted by $H_{\text{bulk}}^{(\text{PBC})}(t)$, satisfies the same commutator relation 
\begin{equation}
   [H_{\text{bulk}}^{(\text{PBC})}(t),H_{\text{bulk}}^{(\text{PBC})}(t^\prime)]=0.
\end{equation}
As a result, the PBC Floquet Hamiltonian is also given by taking a time average: $H_{\text{F}}^{(\text{PBC})}=\frac{1}{T}\int_0^TH_{\text{bulk}}^{(\text{PBC})}(t)\mathrm{d}t$. Under these circumstances, the PBC spectrum can be easily obtained through calculating the eigenvalues of  $h_{\text{F}}(\beta)$ with $\beta$ lying on the conventional Brillouin zone $|\beta|=1$.

Imposing open boundary conditions makes the determination of the quasienergy spectrum of $H_{\mathrm{F}}$ nontrivial. On the one hand, even under the condition in Eq. \eqref{seq:non-Bloch_commutator}, the bulk Hamiltonian $H_{\text{bulk}}(t)$ with open boundaries does not necessarily satisfy $[H_{\text{bulk}}(t),H_{\text{bulk}}(t')]=0$, so that $[H(t),H(t')]\neq 0$. On the other hand, an additional boundary drive $H_{\text{edge}}(t)$ can likewise generate a nonvanishing commutator for $H(t)$, since $[H_{\text{bulk}}(t),H_{\text{edge}}(t')]$ is generically nonzero. Fortunately, under the condition in Eq. \eqref{seq:non-Bloch_commutator}, both situations lead to a commutator $[H(t),H(t')]$ for the full Hamiltonian in Eq. \eqref{seq:H(t)} that is nonzero only near the two boundaries while vanishing in the bulk of the one-dimensional system. 

From the perspective of the evolution operator $U_{\mathrm{F}}$ in Eq.~\eqref{seq:UF}, a time-dependent bulk Hamiltonian satisfying Eq.~\eqref{seq:non-Bloch_commutator} can likewise be understood as generating an effective boundary-driven process. To see this, we define the time-averaged Hamiltonian of $H(t)$ in Eq.~\eqref{seq:H(t)} under open boundary conditions as
\begin{eqnarray}\label{seq:Have}
    H_{\text{ave}}=\frac{1}{T}\int_{0}^{T} H(t)\,\mathrm{d}t.
\end{eqnarray}
Under the condition in Eq.~\eqref{seq:non-Bloch_commutator}, it is straightforward to show that the bulk hopping elements of $H_{\text{ave}}$ are encoded in the Floquet non-Bloch Hamiltonian $h_{\mathrm{F}}(\beta)$ in Eq.~\eqref{seq:h_F_beta}. Since $H_{\text{ave}}$ is time independent and short-ranged, its OBC spectrum can be obtained directly from the conventional non-Bloch band theory for static non-Hermitian systems. One may then define the corresponding evolution operator $U_{\text{ave}}=\exp(-iH_{\text{ave}}T)$, which shares the same eigenstates as $H_{\text{ave}}$.

However, because $[H(t),H(t')]$ remains nonzero near the two boundaries, $U_{\text{ave}}$ does not fully capture the OBC Floquet evolution described by $U_{\mathrm{F}}$. Remarkably, we find that the difference between $U_{\mathrm{F}}$ and $U_{\text{ave}}$ is sharply localized at the boundaries, as shown in Fig.~\ref{sfig1:bulk_matrix} for the representative bulk-driven non-Hermitian model discussed in Sec.~\ref{sec:bulkdriven}. In this sense, the $T$-dependent correction $\Delta U=U_{\mathrm{F}}-U_{\text{ave}}$ can be regarded as an effective boundary drive acting on $U_{\text{ave}}$, thereby inducing resonant coupling between its skin modes once Floquet-zone mixing occurs.

\begin{figure}
    \centering
    \includegraphics[width=0.75\linewidth]{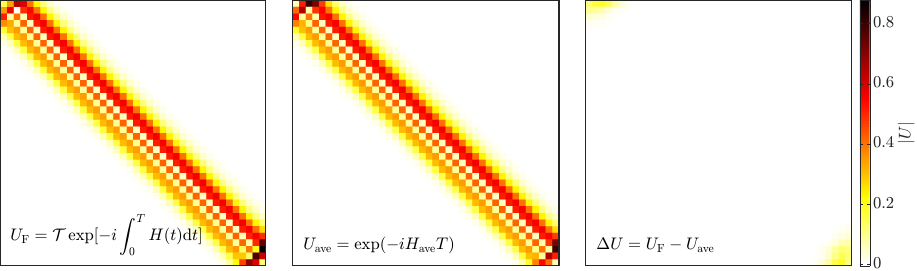}
    \caption{The matrix elements of a Floquet operator $U_{\text{F}}$ (left), an evolution operator $U_{\text{ave}}=\exp(-i H_{\text{ave}}T) $ of a time-averaged Hamiltonian $H_{\text{ave}}=(1/T)\int_{0}^TH(t)\mathrm{d}t$ (middle), and their difference $\Delta U=U_{\text{F}}-U_{\text{ave}}$ (right). These matrix elements are illustrated for an exemplified bulk-driven non-Hermitian system presented in Eq.  \eqref{seq:four_step_Ham} in Sec.  \ref{sec:bulkdriven}, with the parameters $t_1=1$, $t_2=0.08$, $\gamma_1=0.1$, $\gamma_2=0.07$, $T=7.5$, and $L=40$.  }
    \label{sfig1:bulk_matrix}
\end{figure}
As analyzed above, when Floquet-zone folding occurs in the low-frequency regime, the boundary-localized perturbation induced by the nonvanishing boundary commutators can resonantly couple two skin modes and thereby strongly reconstruct the OBC quasienergy spectrum. In this sense, given the short-ranged Floquet non-Bloch Hamiltonian $h_{\mathrm{F}}(\beta)$ in Eq. \eqref{seq:h_F_beta} and the driving period $T$, the Floquet non-Bloch band theory developed in the main text efficiently determines a Floquet generalized Brillouin zone (GBZ) and the corresponding OBC quasienergy spectrum in the thermodynamic limit. 

\subsection{A step-by-step procedure}\label{sec:procedure}
We have already demonstrated the broad applicability of the Floquet non-Bloch band theory developed in the main text. Our framework applies not only to non-Hermitian systems with boundary Floquet drives, but also to a broad class of bulk-driven non-Hermitian systems satisfying Eq. \eqref{seq:non-Bloch_commutator}. In the remaining part of this section, we present a step-by-step procedure for using this theory to extract the OBC quasienergy spectrum in the thermodynamic limit. Whereas the discussion in the main text adopts a geometric viewpoint, tracking the evolution of the Floquet auxiliary generalized Brillouin zones (aGBZs) as the driving period $T$ increases, here we instead fix $T$ and show how to directly calculate the Floquet aGBZs and identify the segments that contribute to the Floquet GBZ.

Under the condition in Eq. \eqref{seq:non-Bloch_commutator}, we start from the Floquet non-Bloch Hamiltonian $h_{\mathrm{F}}(\beta)$ in Eq. \eqref{seq:h_F_beta}. For each $\ell$, the corresponding characteristic equation in Eq. \eqref{seq:characteristic} generally yields $2mq$  roots, denoted by $\beta_1(E+ 2\pi\ell/T), \beta_2(E+ 2\pi\ell/T), \ldots, \beta_{2mq}(E+ 2\pi\ell/T)$. For later convenience, we define $M\equiv mq$ and order these roots in ascending order according to their moduli. 

We first recall that the Floquet aGBZ on the complex $\beta$ plane is defined by the condition that the moduli of two roots from different Floquet zones are equal.:
\begin{eqnarray}\label{seq:floquet_agbz}
\text{aGBZ}_{\ell}\equiv\{\beta: |\beta_i(E)|=|\beta_j(E\pm 2\pi\ell/T)| \},\quad \ell=0,1,2,\cdots.
\end{eqnarray}
Here, $i,j\in\{1,2,\cdots,2M\}$ and the nonnegative integer $\ell=0,1,2,\cdots$ labels the Floquet zone shifted by $\pm2\pi\ell/T$. Specifically, $\text{aGBZ}_0$ is $T$-independent, coincident with the conventional aGBZ of a static non-Hermitian system generated by a time-independent non-Bloch Hamiltonian $h_{\text{F}}(\beta)$ \cite{Yang2020non}.

Before solving the Floquet aGBZ condition in Eq. \eqref{seq:floquet_agbz}, we first clarify the relation between the Floquet aGBZs defined there and the Floquet GBZ that determines the OBC quasienergy spectrum in the thermodynamic limit. As discussed in the main text, the Floquet GBZ can be identified in practice through the following procedure.  We first choose a sufficiently large cutoff $\ell_c$. The required value of $\ell_c$ depends on the extent of Floquet-zone folding induced by a large driving period $T$, and can be determined in practice by verifying that the Floquet GBZ obtained below remains unchanged upon further increasing the cutoff. We then solve the algebraic equations $f(\beta,E+2\pi\ell/T)=0$ for $\ell=-\ell_c,-\ell_c+1,\ldots,\ell_c$, and order all resulting $2(2\ell_c+1)M$ roots according to their moduli as
$|\tilde{\beta}_1(E)|\le|\tilde{\beta}_2(E)|\le\cdots\le|\tilde{\beta}_{2(2\ell_c+1)M}(E)|$.
The Floquet GBZ is then determined by the equality of the moduli of the middle two roots:
\begin{eqnarray}\label{seq:Floquet_GBZ}
    |\tilde{\beta}_{(2\ell_c+1)M}(E)|=|\tilde{\beta}_{(2\ell_c+1)M+1}(E)|.
\end{eqnarray}
 We note that $\tilde\beta_{i}(E)$ with $i=1,2,\cdots, 2(2\ell_c+1)M$ form the same set as $\beta_{j}(E+2\pi\ell/T)$ with $j=1,2,\cdots,2M$ and $\ell=-\ell_c,\ell_c+1,\cdots,\ell_c$. Both of them are the roots of equations $f(\beta,E+2\pi\ell/T)=0$ with $\ell=-\ell_c,\ell_c+1,\cdots,\ell_c$. Therefore, it is easy to see that the Floquet GBZ in Eq.  \eqref{seq:Floquet_GBZ} is a subset of the Floquet aGBZs in Eq.  \eqref{seq:floquet_agbz}. 
 
We now show how to practically solve the Floquet aGBZ condition in Eq. \eqref{seq:floquet_agbz} and identify the Floquet GBZ. This can be achieved using the resultant method introduced in Ref.~\cite{Yang2020non}, which was originally developed to extract aGBZs and the GBZ in static non-Hermitian systems. We note in Eq.  \eqref{seq:floquet_agbz} that, for a fixed $\ell\neq 0$, $\mathrm{aGBZ}_\ell$ consists of two components: $|\beta_i(E)|=|\beta_j(E+2\pi\ell/T)|$ and $|\beta_i(E)|=|\beta_j(E-2\pi\ell/T)|$, where $E\in\mathbb{C}$ is treated without imposing the $2\pi/T$ periodicity on its real part. These two components are equivalent because replacing $E$ with $E+2\pi\ell /T$ transfers the latter into the former. 

We first focus on $|\beta_i(E)|=|\beta_j(E+2\pi\ell/T)|$. It means that, for a specific energy $E$ that solves this equation, there exist a phase variable $\theta\in[0,2\pi)$, such that  $e^{i\theta}\beta_i(E)=\beta_j(E+2\pi\ell/T)$ can be satisfied. Reversely, we can choose a specific $\theta$ and find the common roots $(\beta_c,E_c)$ for the following two equations:
\begin{eqnarray}
\label{seq:two_theta_equation}
\begin{cases}
   f(\beta,E)=0,\\
    f(e^{i\theta}\beta,E+2\pi\ell/T)=0.
    \end{cases}
\end{eqnarray}
It is straightforward to check that the common roots $(\beta_c,E_c)$ of these two equations satisfy $e^{i\theta}\beta_i(E)=\beta_j(E+2\pi\ell/T)$ by identifying $\beta_i(E)=\beta_c$ and $\beta_j(E+2\pi\ell/T)=e^{i\theta}\beta_c$.  Therefore, both $\beta_c$ and $e^{i\theta}\beta_c$ belong to the $\text{aGBZ}_\ell$. The other condition $|\beta_i(E)|=|\beta_j(E-2\pi\ell/T)|$ can be solved similarly by investigating the common roots of two equations $ f(\beta,E)=  f(e^{-i\theta}\beta,E-2\pi\ell/T)=0 $ for a specific $\theta\in[0,2\pi)$. The resulting common roots $(\beta_c^\prime, E_c^\prime)$ have a connection to the above $(\beta_c, E_c)$ via $\beta_c^\prime=e^{i\theta}\beta_c$ and $E_c^\prime=E_c +2 \pi\ell/T$. 

At this stage, the task to locate $\text{aGBZ}_\ell$ is transformed into finding the common roots of two equations in Eq.  \eqref{seq:two_theta_equation} for any $\theta\in[0,2\pi)$. Noting that both $f(\beta,E)$ and $f(e^{i\theta}\beta, E+2\pi\ell/T)$ are finite Laurent polynomials for two variables $\beta$ and $E$, their common roots are thus naturally encoded in the \emph{resultant}, which vanishes when the two polynomials with the same variable have a common root. We first review the definition of the resultant and then show how it facilitates the identification of $\text{aGBZ}_\ell$. Considering two polynomials $p(x)=p_n x^n+\cdots+p_1 x+p_0$   and $q(x)=q_m x^m+\cdots+q_1 x+q_0$, we can define a Sylvester matrix  of dimension $m+n$:
\begin{equation}
\operatorname{Syl}(p, q)=\left(\begin{array}{ccccccc}
p_n & p_{n-1} & p_{n-2} & \ldots & 0 & 0 & 0 \\
0 & p_n & p_{n-1} & \ldots & 0 & 0 & 0 \\
\vdots & \vdots & \vdots & & \vdots & \vdots & \vdots \\
0 & 0 & 0 & \ldots & p_1 & p_0 & 0 \\
0 & 0 & 0 & \ldots & p_2 & p_1 & p_0 \\
q_m & q_{m-1} & q_{m-2} & \ldots & 0 & 0 & 0 \\
0 & q_m & q_{m-1} & \ldots & 0 & 0 & 0 \\
\vdots & \vdots & \vdots & & \vdots & \vdots & \vdots \\
0 & 0 & 0 & \ldots & q_1 & q_0 & 0 \\
0 & 0 & 0 & \ldots & q_2 & q_1 & q_0
\end{array}\right) .
\end{equation}
The resultant of two polynomials $p(x)$ and $q(x)$ is then defined as
\begin{eqnarray}
    \text{Res}_x[p(x),q(x)]=\det\operatorname{Syl}(p, q).
\end{eqnarray}
When there exists at least one common root $x_c$ such that $f(x_c)=g(x_c)=0$, the resultant $\text{Res}_x[p(x),q(x)]=0$.  For example, when $p(x)=x+p_0$ and $q(x)=x+q_0$, the resultant is $\text{Res}_x[p(x),q(x)]=q_0-p_0$, vanishing at $p_0=q_0$ where $p(x)=0$ and $q(x)=0$ have a common root. The function \texttt{resultant(p(x), q(x), x)} is accessible in software programs (e.g., MATLAB), which can be directly used in the numerical investigation.

Therefore, solving the common roots $(\beta_c,E_c)$ for equations in Eq.  \eqref{seq:two_theta_equation} is easily achieved by first finding the resultant of the two polynomials $f(\beta,E)$ and $f(e^{i\theta}\beta,E+2\pi\ell/T)$:
\begin{eqnarray}\label{seq:agbz_resultant}
    g(\beta,\theta)=\text{Res}_E[f(\beta,E),f(e^{i\theta}\beta,E+2\pi\ell/T)],
\end{eqnarray}
where $\beta$ is treated as a parameter entering the coefficients of the polynomials in the variable $E$. The resultant $g(\beta,\theta)$ is therefore a Laurent polynomial in $\beta$, and solving $g(\beta,\theta)=0$ yields $\beta_c(\theta)$ for a fixed $\theta$. We emphasize that $\beta_c$ depends explicitly on $\theta$. Once $\beta_c(\theta)$ is obtained, the corresponding $E_c(\theta)$ can in principle be determined by identifying the common roots of $f(\beta_c(\theta),E)=0$ and $f(e^{i\theta}\beta_c(\theta),E+2\pi\ell/T)=0$. Since our goal is to identify $\mathrm{aGBZ}_\ell$ as the collection of $\beta_c(\theta)$ for $\theta\in[0,2\pi)$, it is sufficient to solve $g(\beta,\theta)=0$. Accordingly, by sweeping $\theta\in[0,2\pi)$ and solving $g(\beta,\theta)=0$ in Eq. \eqref{seq:agbz_resultant}, the collection of all $\{\beta_c(\theta),e^{i\theta}\beta_c(\theta)\}$ gives the complete $\mathrm{aGBZ}_\ell$ defined in Eq. \eqref{seq:floquet_agbz}. This procedure produces the colored Floquet aGBZs shown in Fig.~3 of the main text. Additional examples are presented below.

Once the $\mathrm{aGBZ}_\ell$ with $\ell=0,1,2,\ldots,\ell_c$ have been obtained, the remaining task is to identify which segments of these Floquet aGBZs form the Floquet GBZ that determines the OBC quasienergy spectrum in the thermodynamic limit. This can be carried out as follows. 

Starting from a point $\beta_a$ on the Floquet aGBZs, we first solve $f(\beta_a,E)=0$ to obtain the corresponding set of energies $\{E_{a,1},E_{a,2},\ldots,E_{a,q}\}$. For each $E_{a,i}$, we then solve the family of equations $f(\beta,E_{a,i}+2\pi\ell/T)=0$ with $\ell=-\ell_c,\ldots,\ell_c$, and order all resulting $2(2\ell_c+1)M$ roots according to their moduli as $|\tilde{\beta}_1(E_{a,i})|\le|\tilde{\beta}_2(E_{a,i})|\le\cdots\le|\tilde{\beta}_{2(2\ell_c+1)M}(E_{a,i})|$. We next check the Floquet GBZ condition
$|\tilde{\beta}_{(2\ell_c+1)M}(E_{a,i})|=|\tilde{\beta}_{(2\ell_c+1)M+1}(E_{a,i})|$
in Eq. \eqref{seq:Floquet_GBZ}. Whenever this condition is satisfied, we record the corresponding pair of roots
$\{\tilde{\beta}_{(2\ell_c+1)M}(E_{a,i}),\tilde{\beta}_{(2\ell_c+1)M+1}(E_{a,i})\}$,
which belongs to the Floquet GBZ, together with the associated energy $E_{a,i}$, which contributes to the OBC quasienergy spectrum after resolving the $2\pi/T$ periodicity. Repeating this procedure for all points on the Floquet aGBZs finally yields the Floquet GBZ and the corresponding OBC quasienergy spectrum in the thermodynamic limit.

In summary, we have presented a practical method for determining the Floquet GBZ and the corresponding OBC quasienergy spectrum in the thermodynamic limit, valid for arbitrary driving period $T$. The essential steps are summarized as follows:
\begin{enumerate}
    \item For given model parameters and driving period $T$, choose a sufficiently large integer cutoff $\ell_c$.
    \item Use the resultant method to determine the Floquet aGBZs defined in Eq. \eqref{seq:floquet_agbz}. Concretely, obtain $\mathrm{aGBZ}_\ell$ for $\ell=0,1,\cdots,\ell_c$ by solving Eq. \eqref{seq:agbz_resultant} for different values of $\theta$ and $\ell$.
    \item Identify the segments of the Floquet aGBZs that form the Floquet GBZ by checking whether points on the Floquet aGBZs satisfy the condition in Eq. \eqref{seq:Floquet_GBZ}.
    \item Increase the cutoff from $\ell_c$ to $\ell_c+1$, repeat the above steps, and verify whether the resulting Floquet GBZ remains unchanged. Once this convergence is reached, the Floquet GBZ and the corresponding OBC quasienergy spectrum in the thermodynamic limit are obtained.
\end{enumerate}

Finally, we emphasize that the cutoff $\ell_c$ is introduced solely for practical purposes, namely to account for multiple Floquet-zone foldings in the large-$T$ (low-frequency) regime without having to track an infinite family of characteristic equations, as discussed in the main text. It introduces no approximation. The final Floquet GBZ is a model-determined geometric object that does not depend on the cutoff $\ell_c$. Our framework, therefore, provides an exact method for a broad class of Floquet non-Hermitian systems.

\section{Two additional examples}
\subsection{Boundary-driven two-band model} 

\begin{figure}[t]
    \centering
    \includegraphics[width=0.6\linewidth]{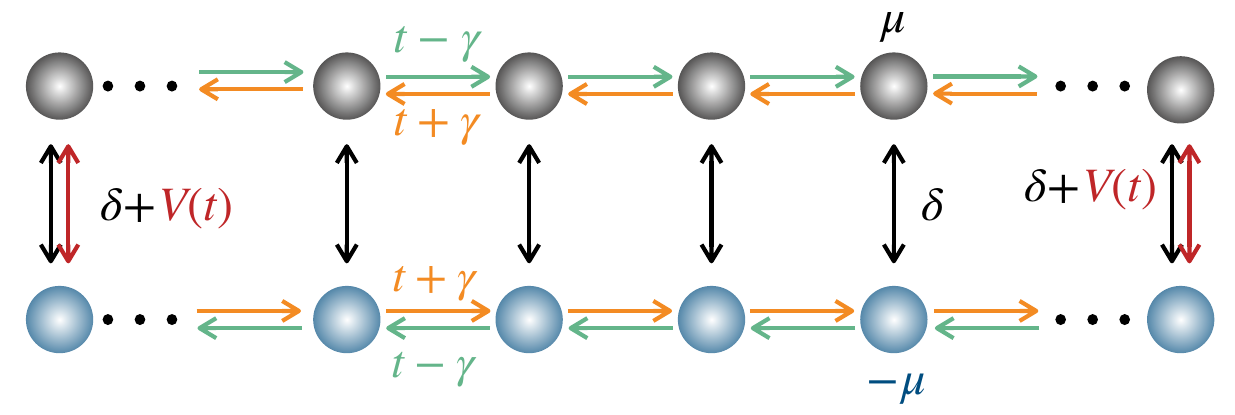}
    \caption{The illustration of a boundary-driven two-band non-Hermitian system. The static bulk Hamiltonian is defined in Eq.  \eqref{seq:twoband_realham}, and the Floquet boundary drive is given by Eq.  \eqref{seq:twoband_boundary}. }
    \label{sfig2:two_band_model}
\end{figure}

In this section, we present an exemplified two-band non-Hermitian system with boundary Floquet drives. As shown below, this model further demonstrates that the idea of boundary Floquet control has a generic applicability in non-Hermitian systems and that the Floquet non-Bloch band theory in Sec.  \ref{sec:theory} applies to  boundary-driven multiband non-Hermitian systems. 

We consider a static bulk Hamiltonian that describes two coupled Hatano-Nelson chains, as shown in Fig. \ref{sfig2:two_band_model}. The time-independent bulk Hamiltonian under open boundary conditions is
\begin{eqnarray}\label{seq:twoband_realham}
    H_{\text{bulk}}=\sum_{j=1}^L[\delta(c_{j,A}^\dagger c_{j,B}+c_{j,B}^\dagger c_{j,A})+\mu(c_{j,A}^\dagger c_{j,A}-c_{j,B}^\dagger c_{j,B})]+\sum_{j=1}^{L-1}\sum_{\alpha\in\{A,B\}}[(t+\xi_\alpha\gamma)c_{j,\alpha}^\dagger c_{j+1,\alpha}+(t-\xi_\alpha\gamma)c_{j+1,\alpha}^\dagger c_{j,\alpha}].
\end{eqnarray}Here, $L$ is the number of unit cells. 
The hopping amplitudes $t\pm\xi_\alpha\gamma$ describe nonreciprocal hoppings within each Hatano--Nelson chain, where $\xi_A=-\xi_B=1$ and $A,\ B$ label the two sublattices. This choice makes the skin modes of the two decoupled chains localize toward opposite boundaries. We remark that the choice of oppositely localized skin modes is made for convenience. The Floquet-induced boundary hybridization generally occurs whenever two skin modes have distinct localization lengths. The two chains, with opposite onsite potentials $\mu_A=-\mu_B=\mu$, are coupled by an interchain hopping $\delta$. Depending on the model parameters, this interchain coupling can strongly reconstruct the OBC energy spectrum and skin modes, giving rise to the critical non-Hermitian skin effect (NHSE) in static non-Hermitian systems~\cite{li2020critical}. The corresponding non-Bloch Hamiltonian is given by
\begin{eqnarray}\label{seq:twobandham}
    h(\beta)=\begin{pmatrix}
       (t+\gamma)\beta+(t-\gamma)\beta^{-1}+ \mu &\delta \\ \delta &  (t-\gamma)\beta+(t+\gamma)\beta^{-1}- \mu
    \end{pmatrix}.
\end{eqnarray}

Here, we are interested in driving this two-band system with a boundary Floquet process. While noting that the microscopic details of the boundary Floquet drives have a negligible influence on the bulk quasienergy spectrum, we concretely consider the following boundary drive protocol in one period:
\begin{eqnarray}\label{seq:twoband_boundary}
    H_{\text{edge}}(t)=\begin{cases}
        +V(c_{1,A}^\dagger c_{1,B}+c_{L,A}^\dagger c_{L,B}+\text{H.c.}),\quad &0\le t<T/2;\\
         -V(c_{1,A}^\dagger c_{1,B}+c_{L,A}^\dagger c_{L,B}+\text{H.c.}),\quad &T/2\le t <T.
    \end{cases}
\end{eqnarray}

With a static bulk Hamiltonian $H_{\text{bulk}}$ and a time-periodic boundary Hamiltonian $H_{\text{edge}}(t)$, this boundary-driven two-band system fits in the framework of Floquet non-Bloch band theory in Sec.  \ref{sec:theory}. Furthermore, as $h(\beta)$ in  \eqref{seq:twobandham} is time-independent, the commuting condition in Eq.  \eqref{seq:non-Bloch_commutator} is automatically satisfied. As a result, a short-range Floquet non-Bloch Hamiltonian in Eq. \eqref{seq:h_F_beta} is just given by $h_{\text{F}}(\beta)=h(\beta)$. Under these circumstances, given a specific driving period $T$, we are able to obtain the Floquet GBZ by following the procedure in Sec.  \ref{sec:procedure}. 
\begin{figure}[t]
    \centering
    \includegraphics[width=\linewidth]{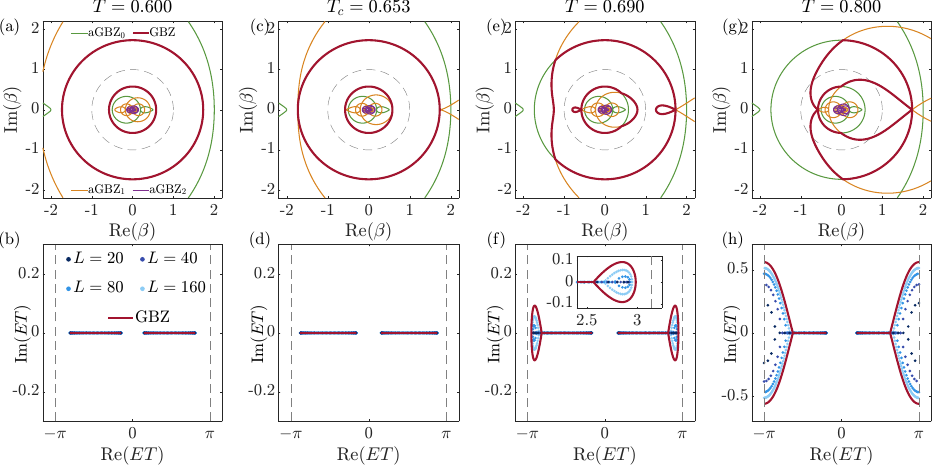}
	\caption{Top row: The Floquet GBZ (red lines) and Floquet aGBZs for the boundary-driven two-band model defined in Eq. \eqref{seq:twoband_realham} and  \eqref{seq:twoband_boundary}. The subscript $\ell$ of $\text{aGBZ}_\ell$ labels the Floquet aGBZ segments contributed from the Floquet zone shifted by $\pm2\pi\ell/T$, defined in Eq.  \eqref{seq:floquet_agbz}. Bottom row: The OBC quasienergy spectra for the boundary-driven two-band model. The colored points are obtained by exactly diagonalizing the OBC Floquet operator $U_{\text{F}}$ for different system sizes. As the system size increases, they converge to the solid lines predicted by Floquet GBZ theory. The different driving periods are labeled at the top of each column. Other parameters are $t=1$, $\gamma= 0.5$,  $\mu=2.5$, $\delta=0.1$, and $V =0.01$.}
    \label{sfig3:two_band_case}
\end{figure}
\begin{figure}
    \centering
    \includegraphics[width=\linewidth]{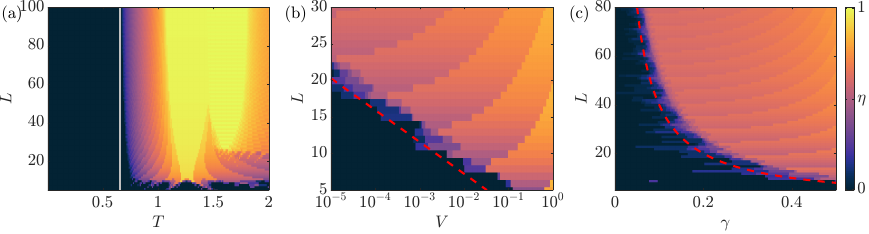}
    \caption{The PT phase diagrams  for the boundary-driven two-band model. The color maps are characterized by the ratio $\eta \in[0,1]$, defined as the fraction of complex eigenvalues in the OBC quasienergy spectrum. (a) The phase diagram of the system size $L$ and driving period $T$, with $\gamma=0.5$ and $V=0.01$. (b) The phase diagram of the system size $L$ and driving strength $V$, with $T=1$ and $\gamma=0.5$. (c) The phase diagram of the system size $L$ and non-Hermitian strength $\gamma$, with $T=1$ and $V=0.01$. In all plots, we set $t=1$, $\mu=2.5$, and $\delta=0.1$. The solid line in (a) is the theoretical prediction $T_c\approx0.653$ from our Floquet GBZ theory, which is shown in Fig.  \ref{sfig3:two_band_case}(c). The red dashed lines indicate the scaling relations $L_c\propto -\log V$ in (b) and $L_c\propto 1/\gamma$ in (c).}
    \label{sfig4:two_band_phase}
\end{figure}

Several representative examples of Floquet GBZs and the corresponding OBC quasienergy spectra in the thermodynamic limit are shown in Fig. \ref{sfig3:two_band_case}. In all cases, the finite-size quasienergy spectra converge toward the thermodynamic-limit results predicted by the Floquet GBZ theory as the system size increases. Before discussing the details, we note that a cutoff $\ell_c=1$ (introduced in Sec. \ref{sec:procedure}) is sufficient for the values of $T$ considered in Fig. \ref{sfig3:two_band_case}. This is evident from the fact that $\mathrm{aGBZ}_2$ does not contribute to the resulting Floquet GBZ in these examples. For larger $T$, however, additional aGBZ components may become relevant, and a practical calculation may therefore require a larger cutoff.

We now analyze the role of the boundary drive in more detail. As discussed in the main text, boundary Floquet driving can resonantly couple two skin modes with distinct localization lengths once the driving period is sufficiently large for Floquet-zone folding to occur. To better isolate this Floquet-engineered effect, we choose a large onsite chemical potential so as to exclude the static critical NHSE reported in Ref.~\cite{li2020critical}. 

With this parameter choice, the OBC quasienergy spectrum remains entirely real in the small-$T$ regime [Fig. \ref{sfig3:two_band_case}(b)], and the corresponding Floquet GBZ is smooth [Fig. \ref{sfig3:two_band_case}(a)]. In this regime, the boundary Floquet driving has only a negligible effect because the large driving frequency cannot resonantly couple the skin modes. As a result, the Floquet GBZ is independent of $T$ and coincides with the static GBZ of the system without boundary driving. 

As the driving period $T$ increases, the $T$-dependent aGBZs move in the complex $\beta$ plane, and $\mathrm{aGBZ}_1$ first touches the Floquet GBZ at a critical period $T_c$ [Fig. \ref{sfig3:two_band_case}(c)]. At this critical point, the quasienergy spectrum remains entirely real. Once $T>T_c$, however, $\mathrm{aGBZ}_1$ starts to contribute to the Floquet GBZ [Figs. \ref{sfig3:two_band_case}(e) and  \ref{sfig3:two_band_case}(g)] and generates several cusps \cite{Hu2024Geometric}. These cusps signal the onset of complex quasienergies [Figs. \ref{sfig3:two_band_case}(f) and  \ref{sfig3:two_band_case}(h)]. The resulting spectral transition indicates the Floquet-engineered non-Bloch parity-time (PT) symmetry breaking, analogous to that of the boundary-driven single-band model discussed in the main text. This PT transition is further illustrated in the phase diagram of Fig. \ref{sfig4:two_band_phase}(a), where the colormap represents the fraction $\eta\in[0,1]$ of complex eigenvalues in the OBC quasienergy spectrum.

We further note that the finite-size quasienergy spectra shown in Fig. \ref{sfig3:two_band_case} exhibit a pronounced system-size dependence. This behavior can again be understood from the perturbative argument presented in the main text. In particular, for a finite system, the boundary Floquet drive in Eq. \eqref{seq:twoband_boundary} becomes significant only above a critical system size $L_c\propto -\log V$, as indicated by the red dashed line in Fig. \ref{sfig4:two_band_phase}(b).  Moreover, we also identify a critical size scale $L_c\propto 1/\gamma$ in the small-$\gamma$ regime [red dashed line in Fig. \ref{sfig4:two_band_phase}(c)]. This scaling is likewise consistent with the perturbative picture in the main text, since for small $\gamma$ the leading-order difference in localization lengths between skin modes is proportional to $\gamma$.

\subsection{Bulk-driven single-band model}\label{sec:bulkdriven}
\begin{figure}
    \centering
    \includegraphics[width=0.5\linewidth]{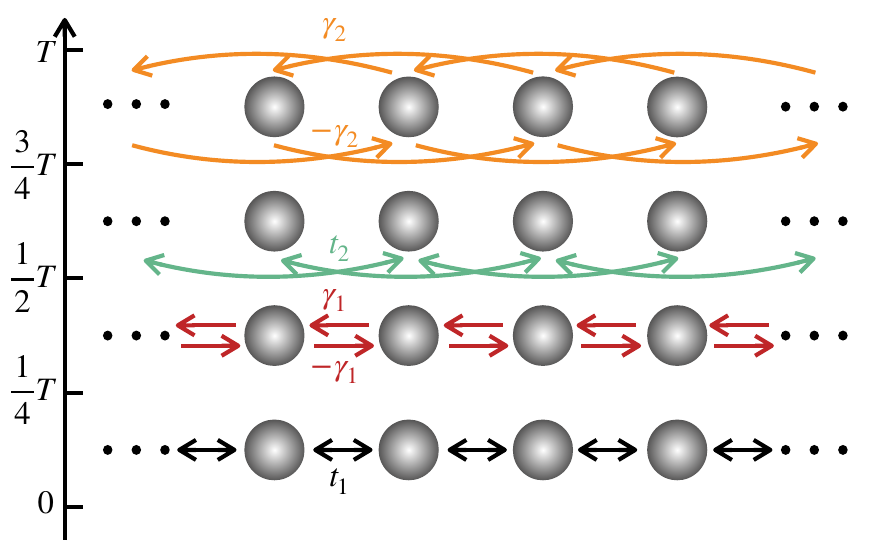}
    \caption{The illustration of a bulk-driven single-band non-Hermitian system, whose time-dependent Hamiltonian under open boundary conditions is defined in Eq.  \eqref{seq:four_step_Ham}.}
    \label{sfig5:bulk_model}
\end{figure}
While the exemplified single-band model in the main text and the above two-band model are driven by boundary Floquet processes, this section presents an example of bulk-driven non-Hermitian systems. This demonstration elucidates that our Floquet non-Bloch band theory developed in Sec.  \ref{sec:theory} also applies to a broad class of bulk-driven non-Hermitian systems. As an illustration, we consider a single-band non-Hermitian chain with a time-periodic bulk Hamiltonian, $H(t)=H(t+T)$, where $T$ is the driving period. There is no boundary Floquet drives, i.e., $H_{\text{edge}}(t)=0$. As shown in Fig.  \ref{sfig5:bulk_model}, $H(t)$ under open boundary conditions contains four driving steps in one period:
\begin{equation}\label{seq:four_step_Ham}
H(t)= \begin{cases}
t_1 \sum_{j=1}^{L-1}c_j^\dagger c_{j+1} +c_{j+1}^\dagger c_j & t \in\left[0,{T}/{4}\right) ,
\\
\gamma_1  \sum_{j=1}^{L-1}c_j^\dagger c_{j+1} -c_{j+1}^\dagger c_j  & t \in\left[{T}/{4},{T}/{2}\right) ,
\\
t_2  \sum_{j=1}^{L-2}c_j^\dagger c_{j+2} +c_{j+2}^\dagger c_j & t \in\left[{T}/{2},{3T}/{4}\right),
\\
\gamma_2  \sum_{j=1}^{L-2}c_j^\dagger c_{j+2} -c_{j+2}^\dagger c_j & t \in\left[{3T}/{4}, T\right).
\end{cases}
\end{equation}
The drive consists of four sequential steps of equal length per period. Two unitary steps implement reciprocal hoppings with amplitudes $t_1$ and $t_2$, and two nonunitary steps implement nonreciprocal hoppings with amplitudes $\gamma_1$ and $\gamma_2$. Remarkably, even though the instantaneous Hamiltonians cannot exhibit NHSE, their combined Floquet evolution gives rise to NHSE.

The corresponding time-dependent non-Bloch Hamiltonian $h(\beta,t)$ is
\begin{equation}\label{seq:four_step_nonblochHam}
h(\beta,t)= \begin{cases}
t_1(\beta+\beta^{-1}) & t \in\left[0,{T}/{4}\right) ,
\\
\gamma_1  (\beta-\beta^{-1})  & t \in\left[{T}/{4},{T}/{2}\right) ,
\\
t_2  (\beta^2+\beta^{-2}) & t \in\left[{T}/{2},{3T}/{4}\right),
\\
\gamma_2  (\beta^2-\beta^{-2}) & t \in\left[{3T}/{4}, T\right).
\end{cases}
\end{equation}

\begin{figure}
    \centering
    \includegraphics[width=\linewidth]{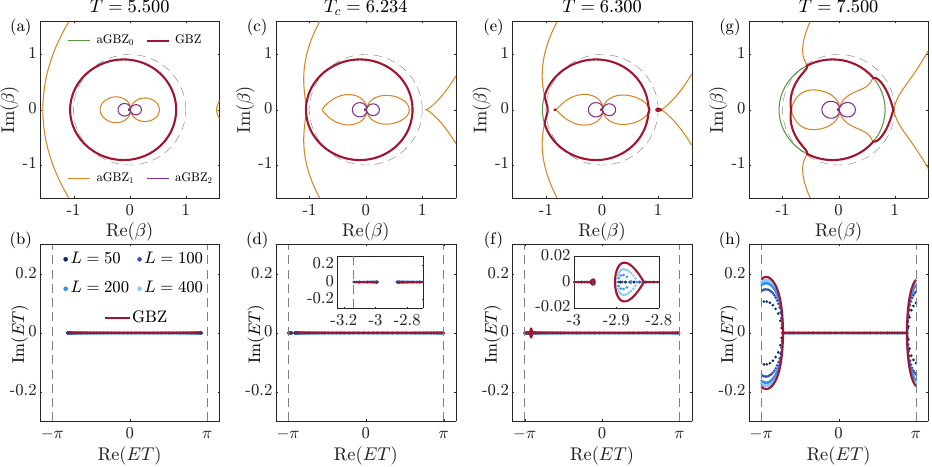}
    \caption{Top row: The Floquet GBZ (red lines) and Floquet aGBZs for the bulk-driven single-band model defined in Eq. \eqref{seq:four_step_Ham}. The subscript $\ell$ of $\text{aGBZ}_\ell$ labels the Floquet aGBZ segments contributed from the Floquet zone shifted by $\pm2\pi\ell/T$, defined in Eq.  \eqref{seq:floquet_agbz}. Bottom row: The OBC quasienergy spectra for the bulk-driven single-band model. The colored points are obtained by exactly diagonalizing the OBC Floquet operator $U_{\text{F}}$ for different system sizes. As the system size increases, they converge to the solid lines predicted by Floquet GBZ theory. The different driving periods are labeled at the top of each column. Other parameters are $t_1=1$, $t_2=0.08$, $\gamma_1=0.1$, and $\gamma_2=0.07$.}
 
    \label{sfig6:bulk_case}
\end{figure}

The time-dependent complex function $h(\beta,t)$ defined above automatically satisfies the condition in Eq.~\eqref{seq:non-Bloch_commutator}. At the same time, the corresponding real-space Hamiltonian $H(t)$ under open boundary conditions has a commutator $[H(t),H(t')]$ that remains nonvanishing near the two boundaries. From the perspective of the Floquet evolution operator $U_{\mathrm{F}}$, this boundary contribution can be viewed as an effective boundary-driven perturbation acting on the evolution operator of an otherwise static system for a time-averaged Hamiltonian [see Fig.~\ref{sfig1:bulk_matrix}]. Therefore, the OBC quasienergy spectrum of this bulk-driven non-Hermitian system can be directly obtained by applying  the Floquet non-Bloch band theory developed in Sec.~\ref{sec:theory}  to the Floquet non-Bloch Hamiltonian:
\begin{eqnarray}\label{seq:non-Bloch_bulk}
        h_{\text{F}}(\beta)=\frac{1}{4}\sum\limits_{n=1}^2(t_n+\gamma_n)\beta^n+( t_n-\gamma_n)\beta^{-n}.
    \end{eqnarray}

\begin{figure}
    \centering
    \includegraphics[width=0.5\linewidth]{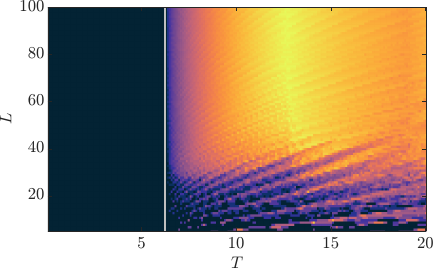}
    \caption{The PT phase diagrams of the system size $L$ and driving period $T$ for the bulk-driven single-band model in Eq. \eqref{seq:four_step_Ham}. The colormap is characterized by the ratio $\eta \in[0,1]$, defined as the fraction of complex eigenvalues in the OBC quasienergy spectrum. The gray solid line represents the critical driving period $T_c$ shown in Fig. \ref{sfig6:bulk_case}(c), which is the location of the Floquet-induced PT transition point. The parameters are $t_1=1$, $t_2=0.08$, $\gamma_1=0.1$, and $\gamma_2=0.07$.  }
    \label{sfig7:bulk_phase}
\end{figure}

Several representative examples of Floquet GBZs and the quasienergy spectra in the thermodynamic limit are presented in Fig. \ref{sfig6:bulk_case}. In all cases, we again find that the finite-size quasienergy spectra converge toward the thermodynamic-limit results predicted by the Floquet GBZ theory as the system size increases. Here, the Floquet GBZ can be efficiently calculated by introducing a cutoff $\ell_c=1$ (see Sec. \ref{sec:procedure}) for the values of $T$ considered in Fig. \ref{sfig6:bulk_case}. This is confirmed by the fact that $\mathrm{aGBZ}_2$ does not contribute to the resulting Floquet GBZ in these examples.

Upon increasing the driving period $T$, the OBC quasienergy spectrum undergoes a real-to-complex transition, signaling PT symmetry breaking. The corresponding phase boundary is set by a critical period $T_c$, at which the Floquet aGBZ component $\mathrm{aGBZ}_1$ first touches the high-frequency Floquet GBZ [see Fig.~\ref{sfig1:bulk_matrix}(c)]. We further present in Fig.~\ref{sfig7:bulk_phase} the PT phase diagram for the fraction $\eta\in[0,1]$ of complex eigenvalues in the OBC quasienergy spectrum. These results demonstrate that the mechanism of Floquet-engineered PT symmetry breaking remains operative in a broad class of bulk-driven non-Hermitian systems.

Finally, we note that, in the large-$T$ regime, the finite-size quasienergy spectra shown in Fig.~\ref{sfig6:bulk_case} exhibit a pronounced system-size dependence. This behavior is reminiscent of the boundary-driven case and can be understood as a Floquet-induced critical NHSE arising from skin-mode hybridization. In contrast to the boundary-driven system studied in the main text, however, the bulk-driven model considered here has an instantaneous Hamiltonian $H(t)$ that exhibits no NHSE at any time $t$. Instead, both the NHSE and its strong finite-size dependence emerge dynamically from interference effects, with the nonvanishing commutator $[H(t),H(t')]$ near the boundaries playing a crucial role in hybridizing the skin modes of the evolution operator $U_{\mathrm{ave}}$ of the time-averaged Hamiltonian. In this sense, our results reveal a new mechanism for dynamically engineering the critical non-Hermitian skin effect. Another important difference from the boundary-driven case is that the effective boundary couplings $\Delta U$ generated by the nonvanishing commutator $[H(t),H(t')]$ near the boundaries remain of the same order as the matrix elements of $U_{\mathrm ave}$ [see Fig.~\ref{sfig1:bulk_matrix}]. As a result, unlike Fig.~\ref{sfig4:two_band_phase}(b) for the boundary-driven system, it is difficult to identify a characteristic length scale in bulk-driven systems.
\bibliography{references}